\documentclass[letterpaper,fleqn,usenatbib]{mnras}

\def\eg{{\it e.g.}}
\def\etal{{\it et al.}}

\def\ie{{\it i.e.}}

\def\vs{{\it vs.}}

\def\Gaia{{\it Gaia\/}}
\def\pmb#1{\setbox0=\hbox{$#1$}%
  \kern-0.25em\copy0\kern-\wd0
  \kern.05em\copy0\kern-\wd0
  \kern-0.025em\raise.0433em\box0}
\def\spmb#1{\setbox1=\hbox{${\scriptstyle #1}$}%
  \kern-0.25em\copy1\kern-\wd1
  \kern.05em\copy1\kern-\wd1
  \kern-0.025em\raise.0433em\box1}

\long\def\Ignore#1{\relax}
\usepackage{color}
\definecolor{red}{rgb}{0.7,0.1,0.1}

\usepackage{graphicx}	

\title[Spiral Scattering]{Discriminating among theories of spiral structure using Gaia DR2}

\author[Sellwood \etal]
          {J. A. Sellwood,$^{1}$\thanks{E-mail:sellwood@as.arizona.edu}
{Wilma H. Trick,$^2$\thanks{E-mail: trick@mpa-garching.mpg.de}}
{R. G. Carlberg,$^3$\thanks{E-mail: raymond.carlberg@utoronto.ca}}
{Johanna Coronado,$^4$\thanks{E-mail:coronado@mpia-hd.mpg.de}}
and
\newauthor {Hans-Walter Rix,$^4$\thanks{E-mail:rix@mpia-hd.mpg.de}}
\\
$^1$Steward Observatory, University of Arizona, 933 N Cherry Ave, Tucson AZ 85722, USA \\
$^2$Max-Planck-Insitut f\"ur Astrophysik, Karl-Schwarzschild-Str. 1, D-85748 Garching b. M\"unchen, Germany \\
$^3$Department of Astronomy and Astrophysics, University of Toronto, Toronto, ON M5S 3H4, Canada \\
$^4$Max-Planck-Insitut f\"ur Astronomie, K\"oningstuhl 17, D-69117 Heidelberg, Germany}

\pubyear{2018}

\begin{document}
\label{firstpage}
\pagerange{\pageref{firstpage}--\pageref{lastpage}}
\maketitle

\begin{abstract}
We compare the distribution in position and velocity of nearby stars
from the \Gaia DR2 radial velocity sample with predictions of current
theories for spirals in disc galaxies.  Although the rich substructure
in velocity space contains the same information, we find it more
revealing to reproject the data into action-angle variables, and we
describe why resonant scattering would be more readily identifiable in
these variables.  We compute the predicted changes to the phase space
density, in multiple different projections, that would be caused by a
simplified isolated spiral pattern, finding widely differing
predictions from each theory.  We conclude that the phase space
structure present in the \Gaia\ data shares many of the qualitative
features expected in the transient spiral mode model.  We argue that
the popular picture of apparently swing-amplified spirals results from
the superposition of a few underlying spiral modes.
\end{abstract}



\section{Introduction}
\label{sec.intro}
The second data release from the \Gaia\ mission \citep{Gaia1} has
revealed a much sharper image of the phase space distribution of
nearby stars \citep{Gaia2}, confirming the existence of multiple
``stellar streams'' passing through the solar neighbourhood.
Follow-up studies \citep[\eg][]{Fama07, Bens07, BH10, Pomp11} of the
smaller star samples in the {\it Hipparcos\/} data, as well as the
GALAH survey \citep{Qu18}, have revealed that each stream contained
stars of a range of abundances and ages, leading to the
well-established conclusion that the phase-space structure was
probably created by dynamical processes within the disc of the Milky
Way.  Dynamical models for some or all of these features have been
presented in many papers \citep[\eg][]{WD00, Quil03, DWT04, QM05,
  Chak07, Anto09, Se10, Gran15, Hu18}.

\citet{TCR} note that the actions contain information about the entire
orbit of the star and not just its instantaneous position and
velocity.  However, \citet{Se10} and \citet{Mc11} showed that the
conjugate angles deduced from the {\it Hipparcos\/} data were not
uniformly distributed, as would be expected if the local stellar
distribution were well mixed, and these additional variables therefore
also contain useful information.  It should be noted that the
transformation from position-velocity space to action-angle variables
assumes first a local model for the radial variation the Galactic
gravitational potential and second that the gravitational potential of
the Galaxy is closely axisymmetric, although both these assumptions
are needed only over the radial range explored by the orbit.  We
describe the evaluation of these variables in \S\ref{sec.vars}.

Our purpose in this paper is to use the \Gaia\ data to
discriminate as far as possible among theories for the origin of
spiral patterns in the discs of galaxies.  Inner Lindblad resonance
(hereafter ILR) features are predicted to be strong in the theoretical
picture presented by \citet{SC14}, and should be weak or absent in
most other theories.  In particular, the ILR should be protected by a
``$Q$-barrier'' in the theory of \citet{BL96}, and the ILR does not
feature prominently in the two other theories: the massive clumps
invoked as drivers by \citet{TK91} and whose non-linear evolution was
studied in simulations by \citet{DOVH} or in theories of continuously
shearing spiral arms \citep[\eg][]{Gran12a, Gran12b, Baba13, Roca13,
  MK18}.

\citet{TCR} highlighted a number of coherent features in the
distribution in action-space that correspond to features seen in the
more usual phase-space, $U$ and $V$.  These features are widely
believed to have been created by dynamical processes.  Here we examine
them further and, in particular, expand the study to include the
additional information contained in the angles conjugate to the
actions.

\section{Sample selection}
\label{sec.select}
\citet{TCR} selected stars from \Gaia\ DR2 \citep{Gaia1} that are
bright enough (G$\;<13$) to have measured line-of-sight velocities.
From their sample, we select stars that have relative uncertainty in
the parallax of $<5\%$ and whose distance from the Sun, projected into
the Galactic plane, is $<200\;$pc.  We also discard stars
having sufficient vertical energy to reach $z>286\;$pc, \ie\ those for
which ${1\over2}[(z\nu)^2 + v_z^2] > 200$, where
$\nu=70\;$km~s$^{-1}$~kpc$^{-1}$ \citep{Fl06}, $z$ is in kpc, $v_z$ is
in km~s$^{-1}$.  This cut limits our sample to stars having thin-disc
kinematics.

\begin{figure}
\includegraphics[width=.98\hsize,angle=0]{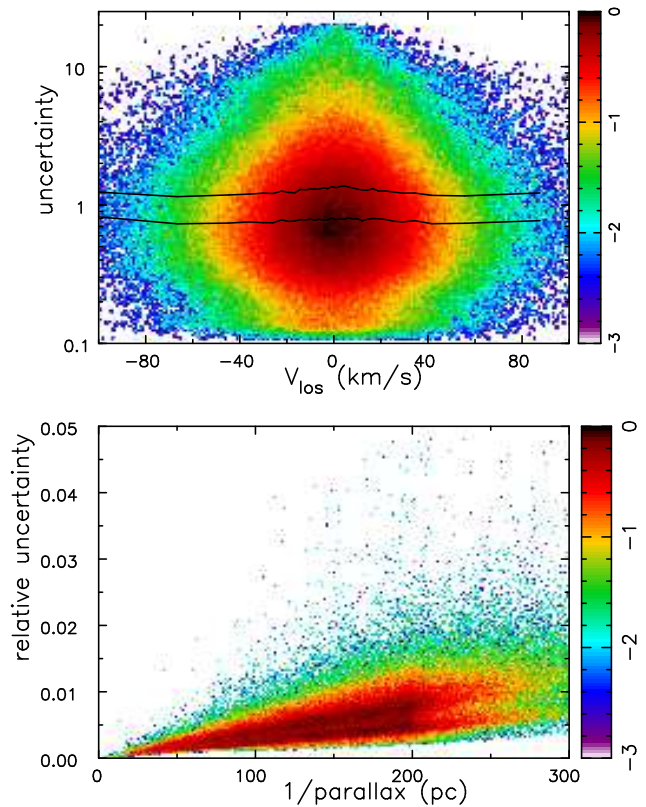} 
\caption{Top panel: The distribution of the line-of-sight velocity and
  its uncertainty for stars in our local sample.  The two near
  horizontal lines mark the mean (upper) and median (lower) of the
  uncertainties in bins of ten thousand stars, indicating that the
  uncertainty is uncorrelated with this componet of velocity.  Bottom
  panel: The distribution of stellar distances and their relative
  uncertainties in the selected sample.  The distance is estimated
  directly from the \Gaia\ parallax, and the uncertainty from their
  quoted value.  In both panels, the logarithmic color scale shows the
  density of stars in the respective coordinates.}
\label{fig.disterrs}
\end{figure}

We further eliminate 3.44\% of the surviving stars that have
uncertainty in the line-of-sight velocity $> 5\;$km~s$^{-1}$.  Note
that the \citet{Gaia2} warn that eliminating stars with larger
velocity uncertainties introduces a bias against high velocity stars,
as indicated in their Figure~7.  However, they consider velocities
resolved into Galactocentric components, which therefore combine
line-of-sight with proper motion components.  In their case, both
transverse components and their uncertainties scale with distance,
leading to their finding of increased uncertainties with larger
velocities.  However, the uncertainty in the line-of-sight component
should not depend on the measured velocity itself, as is confirmed in
the upper panel of Figure~\ref{fig.disterrs}.  Therefore eliminating
stars with large uncertainties in this component only will not
introduce a bias.

These three cuts result in a final sample of $312\,255$ stars, all of
which are sufficiently nearby for the inverse \Gaia\ parallax to yield
a reliable distance.  The lower panel of Figure~\ref{fig.disterrs}
shows that distance uncertainties are better than 1\% for the vast
majority of our stars.

Following \citet{TCR}, we adopt the local standard of rest $(U_\odot,
V_\odot, W_\odot) = (11.1, 12.24, 7.25)\;$km~s$^{-1}$ \citep{SB10},
the Sun's height above the Galactic plane $z_\odot = 25\;$pc
\citep{Ju08}, and Galactic parameters $R_0= 8\;$kpc and $V_0 =
220\;$km~s$^{-1}$ \citep{Bo12}.  From this information, we computed
the full 6D Galactic coordinates, $(R,\phi,z,v_R,v_\phi,v_z)$, for each
of the selected stars.

\section{Action-angle variables}
\label{sec.vars}
For simplicity, we consider motions of stars in the plane of the disc
only and neglect the component of motion normal to the Galactic plane.
Although some interesting features in the vertical motions have been
found \citep[\eg][]{An18, BH18}, most of the in-plane substructure is
found for stars with small vertical action \citep{TCR}.  We therefore
consider only stars whose vertical excursions are confined to
$|z|<286\;$pc.  Also, the rapid vertical oscillations of such stars
should be largely decoupled from their horizontal motion \citep{Se14}.
Thus we need consider just two actions, $J_R$ and $J_\phi \equiv L_z$
and two conjugate angles $w_R$ and $w_\phi$.

The classical integrals, $E = \Phi(R) + {1\over2}(v_R^2 + v_\phi^2)$
and $L_z=Rv_\phi$, can readily be estimated from the \Gaia\ data,
together with some adopted model for the axisymmetric gravitational
potential, $\Phi(R)$, in the disc mid-plane.  We compute the radial
action from the approximate expression
\begin{equation}
J_R = {1 \over \pi} \int_{R_p}^{R_a} \dot R dR \quad\hbox{with}\quad
\dot R^2 = 2\left[E - \Phi(R)\right] - {L_z^2 \over R^2},
\label{eq.radact}
\end{equation}
where $R_p$ and $R_a$ are respectively the Galactic peri- and
apo-centric distances of the star where its radial speed $\dot R =0$;
the value is normalized by $\pi$ because the integral covers only half
a full radial oscillation.  This expression yields very nearly the
same value \citep{Mc11} for $J_R$ as obtained from the more-accurate
``torus fitting'' method that takes account of 3D motion.  Physically,
the radial action is an angular momentum-like variable that quantifies
the magnitude of the radial motion of a star and, for disc stars, it
is typically an order of magnitude smaller than the orbital angular
momentum.

\begin{figure}
\includegraphics[width=.64\hsize,angle=270]{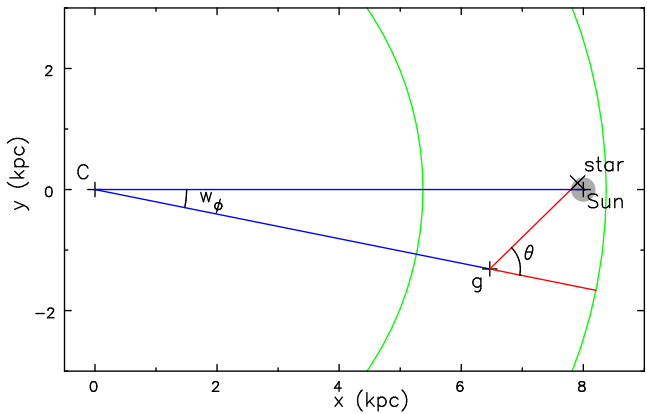} 
\caption{Sketch showing part of the Milky Way seen in projection; all
  the stars in our sample are confined to the lightly shaded region
  around the Sun.  The Sun moves in clockwise direction about the
  Galactic centre, which is marked as C.  A star in our sample might
  be located at the position shown by the X, and moves on a eccentric
  orbit between the radii marked with green circles at $R_p$ and
  $R_a$.  The angle $w_\phi (> 0)$ is the Galactic azimuth of the
  instantaneous position of the guiding centre, which is marked g,
  and is reckoned from the line from the Sun to C.  The angle $w_R$,
  which is conjugate to the radial action, $J_R$, gives the phase of
  the star around its epicycle, but is {\it not exactly} equal to
  $\theta$, and is instead determined as described in the text.  Note
  that $w_R>0$ for a star at the indicated position because stars move
  in a retrograde direction about their guiding centres.}
\label{fig.drawing}
\end{figure}

Note that expression (\ref{eq.radact}) for $J_R$, and the evaluation
of the angle variables described below, require multiple orbit
integrations for each star.  Each individual quadrature is fast, so
that the computational cost of evaluating them for all $312\,255$
stars in our sample is not particularly burdensome.  However, the
implicit relations between the angles and the peculiar velocities that
we need to invert repeatedly require many millions more evaluations,
making it computationally advantageous to choose the simple functional
form, $\Phi(R) = V_0^2\log(R/R_0)$ implying an exactly flat rotation
curve, for the local radial variation of the Galactic gravitational
potential.  We have preferred this form over the Galaxy model proposed
by \citet{Bo15}, even though Bovy's coordinate conversions are
effected quite efficiently using a local ``St\"ackel fudge''
\citep{Bi12}.  We have verified that the distribution of values of the
actions and angles is closely similar whichever of these two
assumptions is adopted, and the features in the distributions of
action-angle variables are insensitive to the chosen model.

Figure~\ref{fig.drawing} illustrates the meaning of the angle
$w_\phi$, which is the Galactic azimuth of the star's guiding centre
and we choose $w_\phi=0$ for the line from the Galactic centre that
passes through the Sun.  The angle $w_R$ is the phase of the star
around its epicycle, and we choose $w_R=0$ at apocentre.  These angles
are computed from the positions reported by \Gaia, but both increase
with time at the uniform rates $\dot w_R = \Omega_R$ and $\dot w_\phi
= \Omega_\phi \equiv L_z/R_{\rm g}^2$, where $R_{\rm g}$ is the
Galactic radius of the star's guiding centre.  The period of the
radial oscillation defines the uniform rate
\begin{equation}
\Omega_R = {2\pi \over \tau_R} \qquad\hbox{where}\qquad \tau_R =
2\int_{R_p}^{R_a} \dot R^{-1} dR.
\end{equation}
Since $w_R$ increases uniformly with time, it does not correspond
precisely to the angular position of the star around its epicycle,
marked $\theta$ in the sketch, because the radial motion is generally
not perfectly harmonic.  To evaluate $w_R$ for a star at a general
point, we must integrate its orbit to find the time $t$ required for
it to reach its present radius from the moment it passed through
apocentre.  Then $w_R = \Omega_R t$.  It is always true that
$w_R=\pm\pi$ exactly at pericentre and, because epicycle motion is
slower in the outer part, $|w_R| \ga\pi/2$ as the star crosses the
circle at $R_{\rm g}$ in respectively the inward and outward
directions.

This discussion is valid for orbits of arbitrary eccentricity, but the
radial motion becomes more nearly harmonic as the orbit eccentricity
decreases, and the angular frequencies tend to the familiar
definitions introduced long ago by Lindblad \citep[\eg][]{BT08}:
\begin{equation}
\Omega_\phi \rightarrow \Omega_c \qquad\hbox{and}\qquad
\Omega_R \rightarrow \kappa,
\end{equation}
where $R\Omega_c^2 = d\Phi/dR$ and $\kappa^2 = 4\Omega_c^2 + R d\Omega_c^2/dR$.

We show that the mapping from the observed 4D Galactic phase-space
coordinates of a star at $(R,\phi,v_R,v_\phi)$ to
$(J_R,L_z,w_R,w_\phi)$ is well-behaved, and \citep[as was stressed
  by][]{TCR} any features present in the one projection must
necessarily appear in any other, although in a distorted form.  In
this paper we also demonstrate that action-angle coordinates have the
additional advantage that they clarify the connection between the
substructure and its probable dynamical origin.

\section{Features in phase space}
\label{sec.slopes}
\begin{figure}
\includegraphics[width=.97\hsize,angle=0]{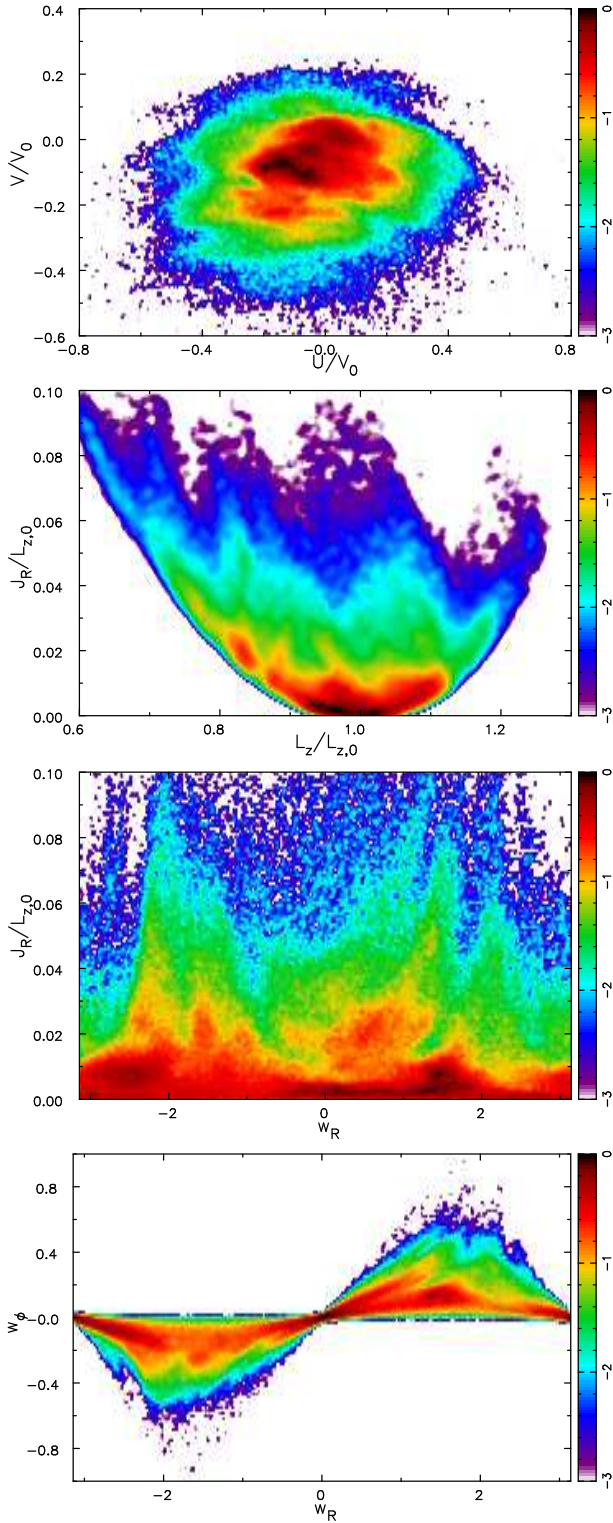} 
\caption{Four projections of the phase space density of our sample of
  nearby stars that stay close to the disk plane selected from \Gaia
  DR2.  The top panel shows the usual velocity-space $(U,V)$
  distribution, the second panel the distribution in the space of the
  two actions $(L_z,J_R)$, the third panel the distribution in the
  space of the radial angle and action $(w_R,J_R)$, and the bottom
  panel the distribution in the space of the two angles
  $(w_R,w_\phi)$.  The scales of $L_z$ and $J_R$ have been normalized
  by the angular momentum of the LSR.  The color scale represents the
  logarithm of the relative density in each projection.}
\label{fig.Gaia_all}
\end{figure}

Figure~\ref{fig.Gaia_all} presents the phase space density of our star
sample in four different projections.  The top panel reproduces the
helio-centric velocity-space $(U,V)$ distribution \citep{Gaia2}, which
manifests multiple features that have become known as streams.  The
stars are projected into the space of the two actions, $(L_z,J_R)$, in
the second panel, which was also presented by \citet{TCR}, where both
actions are normalized by the angular momentum of the LSR.  The
parabolic lower boundary to this distribution results from our
selection of stars that are today passing within $200\;$pc of the Sun,
since those stars with $L_z \neq L_{z,0}$ will pass close to the Sun
only if their orbits are sufficiently eccentric, \ie\ the greater the
value $|L_z - L_{z,0}|$, the larger $J_R$ must be for the star to be
in our sample.  The distribution in $(w_R,J_R)$ space is illustrated
in the third panel, while the bottom panel shows the distribution in
$(w_R,w_\phi)$ space.

\begin{figure}
\begin{center}
\includegraphics[width=.91\hsize,angle=0]{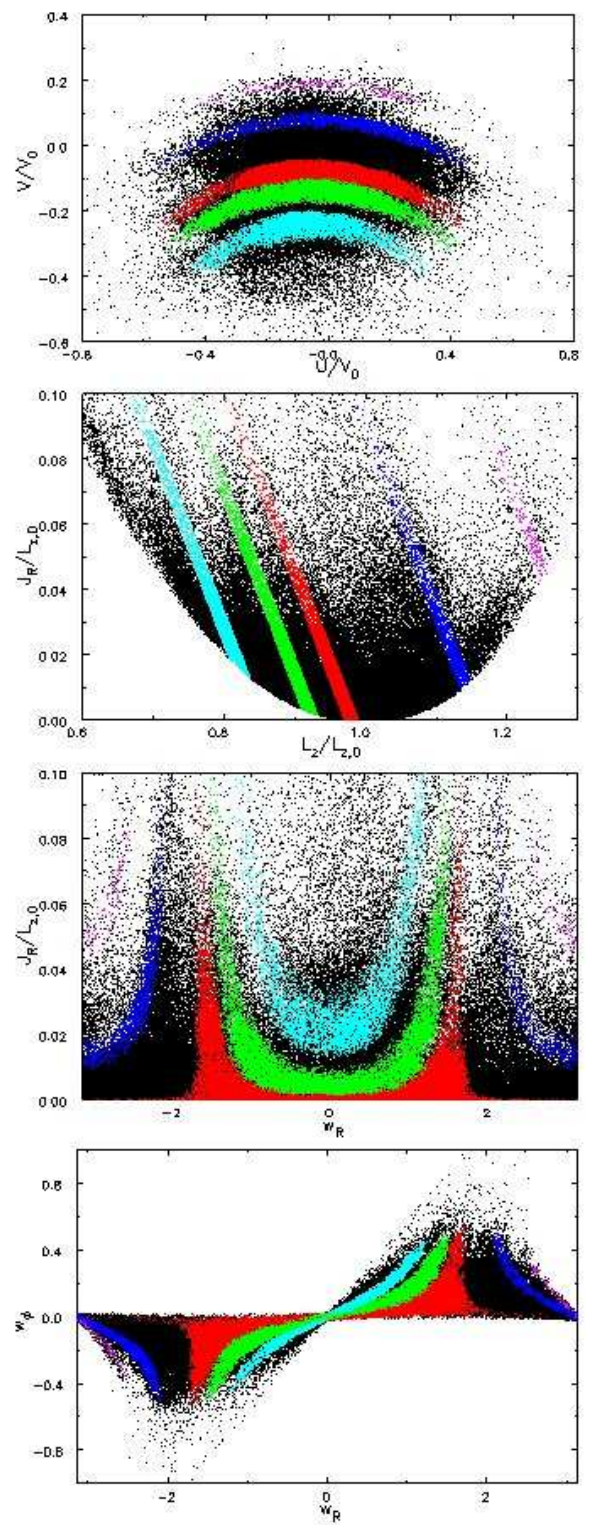}
\end{center}
\caption{As in Fig~\ref{fig.Gaia_all}, but showing the individual
  \Gaia DR2 stars.  The coloured stars were selected to highlight five
  separate sloping features in the second panel, and the other panels
  reveal where the same stars lie in the separate projections.  We
  chose the centers of the sloping lines to intercept the $x$-axis at
  0.85 (cyan), 0.925 (green), 0.98 (red), 1.16 (blue), and 1.33
  (magenta).}
\label{fig.mult_res}
\end{figure}

Figure~\ref{fig.mult_res} illustrates how stars are mapped between
these different coordinate projections.  We have highlighted stars in
five distinct groupings by colour, each coloured group being those
stars lying within $\Delta L_z=0.01$ of a line of slope $-0.5$ in the
second panel, for reasons that will become clear later.  The
intercept on the $L_z$ axis of each line was chosen by eye so that
each selected group of stars included one of the more prominent
features in that panel.  The coloured points in the other panels of
this Figure indicate where each group of selected stars lies in the
three other projections.  While the mappings from one panel to the
other are complicated, they are well-behaved, and the highlighted
stars lie in overdense regions in each projection.

\begin{figure}
\includegraphics[width=\hsize,angle=0]{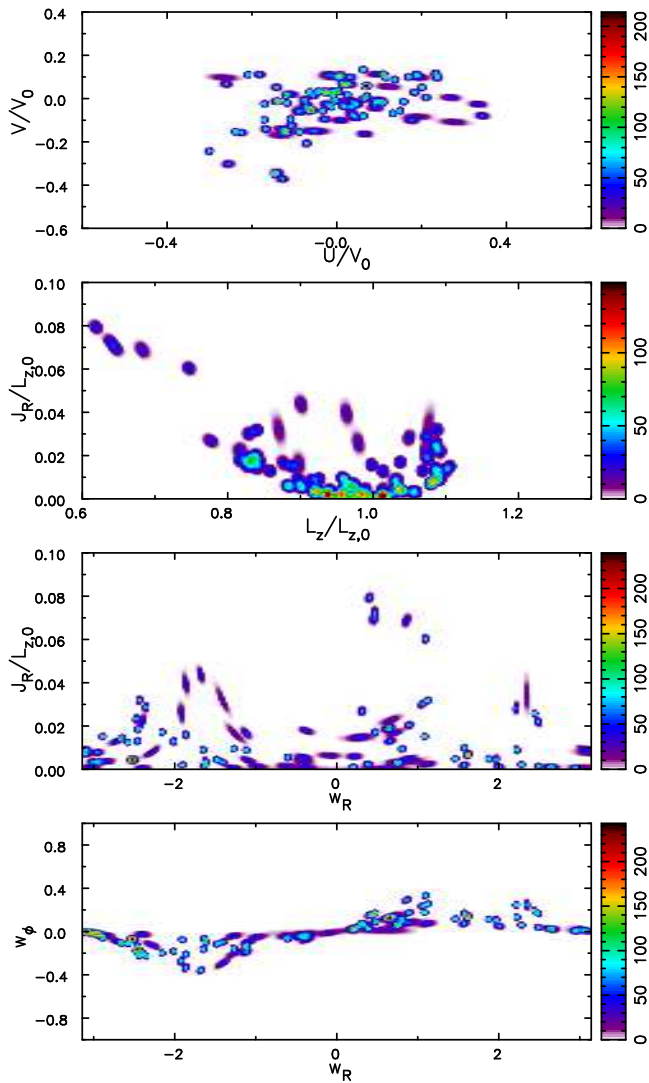} 
\caption{Estimated uncertainties in the quantities of interest here,
  derived for the first 100 stars in our sample.  Each smudge shows
  the spread in values, coloured on a linear scale, that results from
  1000 Monte Carlo realizations, using the uncertainties in each
  observed quantity, as reported for each star in \Gaia DR2.}
\label{fig.uncerts}
\end{figure}

Figure~\ref{fig.uncerts} shows uncertainties in the derived quantities
for the first 100 stars in our sample.  For each star, we drew one
thousand sets of new values for RA, dec, parallax, proper motions in
RA and in dec, and radial velocity, each drawn from a Gaussian
distribution with the observed value as the assumed mean and the
\Gaia\ estimate of the uncertainty in each quantity as the standard
deviation.  We used each realization of these adjusted coordinates to
derive revised estimates of the velocities, actions and angles, and
show the density of these new values in each projection.  From this
Figure, we see that the uncertainties in the derived quantities are
generally small, but are somewhat larger for several stars, although
the uncertainties, naturally, remain small compared with the widths of
the features in Figure~\ref{fig.Gaia_all}.  The principal source of
uncertainty appears to arise from the line-of-sight velocity, even
though we discarded stars having uncertainties in this component $>
5\;$km~s$^{-1}$.

\begin{figure}
\includegraphics[width=.5\hsize,angle=270]{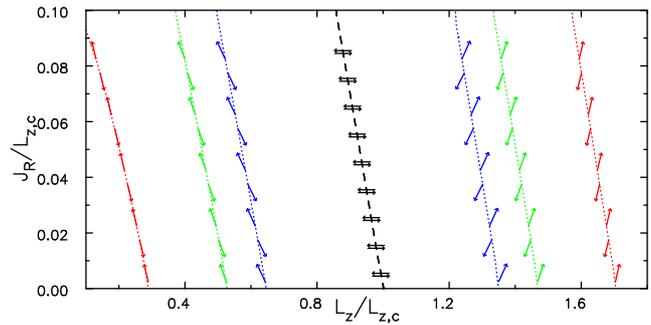} 
\caption{Scattering in action space for a self-similar $V=\;$const
  galaxy model.  Both actions are normalized by $L_{z,c}$, which is
  the angular momentum of a circular orbit at corotation.  The broken
  lines mark the separate loci of CR (black) and Lindblad resonances,
  colored red for $m=2$, green for $m=3$ and blue for $m=4$, computed
  using eq.~(\ref{eq.resonance}) for orbits of finite eccentricity.
  The vectors, having arbitrarily chosen lengths, indicate possible
  changes computed from eq.~(\ref{eq.Jacobi}).
  Formula~(\ref{eq.actions}) indicates that they should have slopes
  $l/m$.}
\label{fig.Jacobi}
\end{figure}

\subsection{Features in the action distribution}
The substructure in the second panel of Figure~\ref{fig.Gaia_all} may
be indicative of resonance scattering.  Here we explain why we formed
this view.

A test particle moving in a non-axisymmetric potential that rotates at
the uniform rate $\Omega_p$ about the $z$-axis conserves neither its
energy $E$ nor its angular momentum $L_z$, but Jacobi's integral $E_J
\equiv E - \Omega_pL_z$ is conserved \citep{BT08}.  Hence, any changes
to the energy and angular momentum of a star must be related as
\begin{equation}
\Delta E = \Omega_p \Delta L_z.
\label{eq.Jacobi}
\end{equation}
When the non-axisymmetric part of the potential is weak and has
$m$-fold rotational symmetry, the motion of a star resonates with
the rotating potential whenever
\begin{equation}
m(\Omega_p - \Omega_\phi) = l\Omega_R,
\label{eq.resonance}
\end{equation}
with $l=0$ for the corotation resonance (CR) and $l=\pm1$ for the
Lindblad resonances.  At the ILR the star overtakes the wave, while at
the OLR it is overtaken by the wave, and in both cases it experiences
forcing at its natural radial frequency $\Omega_R$.  \citet{LBK}
showed that only those stars that are in one of these resonances
experience lasting changes to $E$ and $L_z$.  The broken curves in
Figure~\ref{fig.Jacobi} mark the loci of the resonances in action
space for $l=0,\; \pm 1$ and $m=2$ (red), $m=3$ (green), and $m=4$
(blue) perturbations in a simple $V=\;$constant galactic potential.
These curves, which are calculated exactly for eccentric orbits using
eq.~(\ref{eq.resonance}), have similar negative slopes in any
reasonable galaxy potential.

\citet{SB02}, in just a few lines of algebra, were able to show that
scattering at any one of these resonances required not only that
$\Delta E = \Omega_p \Delta L_z$, but that changes to the actions were
simply related as
\begin{equation}
\Delta J_R = {l\over m}\Delta L_z.
\label{eq.actions}
\end{equation}
Formally, this relation is for near circular orbits, but the predicted
slope holds for quite eccentric orbits, as illustrated by the vectors
in Figure~\ref{fig.Jacobi}, with $\Delta J_R$ computed exactly from
eq.~(\ref{eq.Jacobi}) and not the approximation (\ref{eq.actions}).

The horizontal vectors indicate that $\Delta J_R = 0$ for moderate
$\Delta L_z$ changes at CR where $l=0$, with the implication that
radial migration at this resonance occurs without increasing random
motion.

However, scattering at either Lindblad resonance should be along
trajectories of almost constant slope $l/m$ in $L_z$-$J_R$ space,
which therefore have positive slope at the OLR and negative at the
ILR.\Ignore{ The slopes of the vectors are shallower for disturbances
  of higher rotational symmetry, \ie\ larger $m$.}  \citet{LBK} showed
that stars may either gain or lose $L_z$ at all the major resonances,
as indicated by the vectors in Figure~\ref{fig.Jacobi}, but they also
showed that on average stars lose $L_z$ at the ILR and gain at the
OLR, so formula (\ref{eq.actions}) predicts net heating because the
average $J_R$ is increased at both resonances.  Notice that the
scattering vectors for all initial $J_R$ values shift stars off the
OLR as they gain or lose $L_z$, but the scattering vectors at the ILR
are almost perfectly aligned with the resonance line for $m=2$ (red)
and misalignment develops slowly as $m$ increases.  This near
coincidence is not just a pecularity of this potential and appears to
hold in most other reasonable galactic potentials.  Thus stars at the
ILR remain close to the resonance as they are scattered, which allows
large changes in $J_R$ to be built up, as we will demonstrate.

In summary, the second panel of Figure~\ref{fig.Gaia_all} manifests
multiple density excesses having predominantly negative slopes that
are similar to the slopes $\approx -0.5$ of all the major resonances
in Figure~\ref{fig.Jacobi}.  These features therefore suggest, but do
not prove, that the stellar distribution in the neighbourhood of the
Sun contains many stars that have been scattered at Lindblad
resonances of several different disturbances having various pattern
speeds.

It should now be clear why we chose to highlight stars along lines of
slope $-0.5$ in Figure~\ref{fig.mult_res}, since they could be stars
that have been scattered at resonances.  Whether the density excesses
are or not created by resonant scattering, the highlighted stars in
other panels show the loci of possible resonances in the other
projections.

\subsection{Features in the angle distribution}
The third panel of Figure~\ref{fig.Gaia_all} shows the distribution of
the \Gaia\ stars as a function of $w_R$ and $J_R$.  As for all other
projections of the 4D distribution, we see a rich substructure.  Our
attempts to reproduce features in the different projections are
described in \S\ref{sec.predictions}, where the possible origin of the
features in this projection will become clear.

The narrow distribution of $w_\phi$ values in the bottom panel, is a
consequence of our selection of stars close to the Sun.  We see that
stars near apocentre ($w_R=0$) and pericentre ($w_R=\pm\pi$) all have
$w_\phi\approx 0$, as expected.  The largest values of $w_\phi$ arise
for stars having eccentric orbits that are close to their guiding
centre radii ($w_R \ga \pm\pi/2$), which is consistent with the sketch
in Figure~\ref{fig.drawing}.  Because the $w_\phi$ distribution is so
dominated by selection effects, we have not found this last projection
very informative.

By comparing the three other projections of the \Gaia\ data with
models in the next section, we will argue that the current stellar
distribution has been sculpted by multiple resonances over time.

\section{Predictions of various theories of spiral structure}
\label{sec.predictions}
Here we calculate how an originally smooth distribution of stars in an
axisymmetric potential model is changed when perturbed by various
time-dependent, non-axisymmetric perturbations.  Our aim is not to
create a single model that can match all the features in the Gaia
data, but more modestly to show qualitatively how some of features
might have arisen. 

Each of our perturbations is motivated by one of the current theories
for the origin of spirals in galaxies.  \citet{SC14} proposed that
spirals result from transient modes, which we descibe more fully and
test in \S\ref{sec.mode}.  A number of authors have proposed that
spirals are material arms, and we examine the consequences of this
hypothesis in \S\ref{sec.materialarm}.  \citet{TK91} and \citet{DOVH}
imagine that spirals result from the responses to mass clumps within
the disk, which we model in \S\ref{sec.massclump}.  Finally,
\citet{BL96} propose that spirals are long-lived, quasi-steady density
waves, and we discuss their picture in \S\ref{sec.qssh}.

\subsection{Method}
\label{sec.method}
We employ the method pioneered by \citet{WD00}, which assumes some
reasonable distribution function (DF) for the Galaxy before the
perturbation was introduced and uses the fact that the DF is conserved
along any orbit from the moment a perturbing potential is added to the
present day.  \citet{Bo15} also implemented the technique in the python
package \texttt{galpy}.  Since we need to know the DF only where we wish to
compare it with data, its current value at any point $(R, \phi, v_R,
v_\phi)$ can be obtained by integrating an orbit from that point
backwards in time over the history of the perturbation to when the
adopted model was smooth and axisymmetric, where the value of the DF
can be obtained from its coordinates $(R^\prime, \phi^\prime,
v_R^\prime, v_\phi^\prime)$ at that earlier moment.

We adopt this method to model phase space at the location of the Sun.
For each point in $(L_z,J_R)$ space, say, we must determine what the
values of those variables imply for the instantaneous velocities of a
star that also passes through the position of the Sun.  That is we
must find the $(R,\phi,v_R,v_\phi)$ coordinates of that orbit for
fixed $R=R_0$ and $\phi=0$.  Trivially, $v_\phi = L_z/R_0$, and we map
$(L_z,J_R) \rightarrow (E,L_z)$ and then use $v_R^2 = 2[E-\Phi(R_0)] -
(L_z/R_0)^2$.  We then integrate back in time to find where the orbit
was before the perturbation, to find the value of the DF at the
starting point in phase space.  Note the two possible signs of $v_R$
imply that two separate locations $(R^\prime, \phi^\prime, v_r^\prime,
v_\phi^\prime)$ before the perturbation was introduced map to the same
point in $(L_z,J_R)$ space, and we therefore sum both DF values.  The
calculation when $w_R$ is pre-specified is more difficult, since we
need a numerical search over $(E,L_z)$ for the unique orbit that
passes through $(R,\phi) = (R_0,0)$ with phase $w_R$ and radial action
$J_R$; once found, the $v_\phi$ and $v_R$ values are determined as
before, but the value of $w_R$ removes the sign ambiguity of $v_R$.

We calculate the perturbed DF at the point $(R,\phi) = (R_0,0)$, and
make no attempt to average over the small area of radius 200~pc that
contains the stars that we compare with our models.  Our selection of
such nearby stars was partly motivated by the need to be able to
neglect gradients in phase space across this area.

We assume a ${\rm sech}^2\gamma t/2$ time dependence for every
perturbation, which asymptotes to exponential growth as $\exp(-\gamma
t)$ for $\gamma t \ll -1$ and begins to decay after peaking at $t =
0$.  Typically, the value of $\gamma = 0.1\Omega_p$.  We begin the
integration at $t=1.36\;$Gyr, which is six orbit periods at the Sun
($2\pi R_0/V_0$) after the perturbation peaked at $t=0$, and integrate
backwards in time to when the perturbation had negligible amplitude.
The choice of six orbit periods after saturation allows the
perturbation to decay somewhat, but is before some of the perturbed
features become affected by a slow phase wrapping.  We justify this on
the grounds that continued evolution of a single perturbation would
not be observable in the \Gaia\ data because all theories, bar one,
predict successive perturbations on a time-scale of a few orbits.  The
exception is the quasi-steady mode theory of \citet{BL96} that we
argue (\S\ref{sec.qssh}) predicts no phase space changes at all.

Our unperturbed Milky Way model is locally approximated as a disc
having a simple Gaussian velocity distribution, with radial velocity
dispersion $\sigma_R= 0.13V_0$.  The precise degree of random motion
in the disc not very important, since we are here concerned with the
qualitative relative changes produced by each perturbation, and do not
attempt a quantitative comparison.  The DF is
\begin{equation}
f(E,L_z) \propto \exp(-1.5R_g) e^{-{\cal E}/\sigma_R^2},
\end{equation}
where $R_g = L_z/V_0$ is the guiding center radius and ${\cal E} = E -
E_c(L_z)$, is the excess energy above that of a circular orbit at
$R_g$; $E_c = \Phi(R_g) + V_0^2/2$.  The $\exp(-1.5R_g)$ factor was
chosen to yield about the right variation with $L_z$.  We ``observe''
the perturbed DF that results from each adopted perturbation from the
location of the Sun at $R_0 = 8\;$kpc and $\phi=0$, and in the
following figures we normalize the phase space density by its maximum
value in every projection.

\begin{figure}
\includegraphics[width=\hsize]{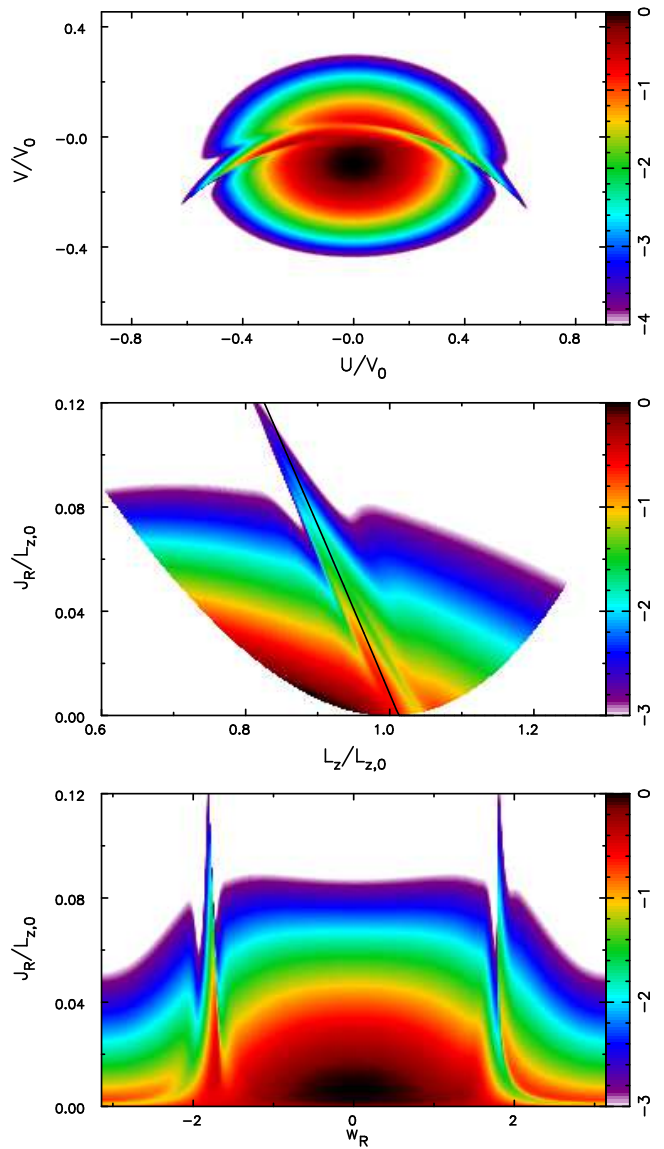} 
\caption{The predicted phase-space density of stars in three
  projections after our model spiral mode, described in the text, has
  grown and started to decay.  The DF is largely undisturbed, except
  for features resulting from ILR scattering, that are broadened by
  the time dependence of the perturbing potential.  The locus of the
  ILR is marked by the line in the middle panel.  As above, the color
  scale represents the logarithm of the relative density in each
  projection.}
\label{fig.spiral_ILR}
\end{figure}

\subsection{A transient spiral mode}
\label{sec.mode}
\citet{SC14} proposed that spirals result from relatively short-lived
unstable modes.  Specifically, this description means that each
perturbation is a standing wave oscillation, having a constant shape
that rotates at a steady rate and grows exponentially until it
saturates.  Their simulations revealed that each coherent mode had a
lifetime at moderate amplitude of some 10 rotation periods at its
corotation radius, and the disk generally supported a few such modes
simultaneously, with new instabilities developing in constant
succession.  The superposition of several coherent modes gave rise to
the continuously changing patterns that are observed in most
simulations.  While the classical density wave theory \citep{BL96}
also invokes modes, those authors expect each mode to be very slowly
growing, or even ``quasi-steady'', and the disk would support rather
few such modes; \ie\ they have a very different picture from that
proposed by \citet{SC14}.

In order to calculate what to expect from a single transient spiral of
the kind described by \citet{SC14}, we chose a 3-armed spiral
perturbation with a pattern speed $\Omega_p =
14.4\;$km~s$^{-1}$~kpc$^{-1}$, for which the ILR lies at 8.1~kpc.  We
adopted a logarithmic spiral potential that had a pitch angle of
$26^\circ$ and a radial variation that peaked at corotation and
decreased as a quartic polynomial to zero at a radius that was 80\% of
that of the ILR -- \ie\ it was weak, but non-zero, a little interior
to the ILR.  As described above, it had a ${\rm sech}^2$ time
dependence, with the asymptotic growth rate $\gamma = 0.1\Omega_p$.
We believe these choices to be reasonable.  The scattering plots we
present below were little changed when we employed a more tightly
wrapped spiral, and differed only slightly for $m=2$ or $m=4$ patterns
because of the slight slope changes to the scattering vectors in
Figure~\ref{fig.Jacobi}.  We have also experimented with wide
variations in the pattern speed as we report below.

Although \citet{SC14} identified growing modes, and showed that the
frequency persisted for some time after saturation, the assumption we
have made here that the shape remains the same as the disturbance
decays is convenient, but probably too simple.

Figure~\ref{fig.spiral_ILR} shows three projections of the phase space
density of stars for this first case. The DF is largely undisturbed in
all three panels except for narrow features that have been caused by
ILR scattering, and each feature in these noise-free projections
illustrates the bi-directional nature of the scattering expected from
Figure~\ref{fig.Jacobi}.  Both actions are normalized by the $L_z$ of
the LSR in the middle panel, and the requirement that the stars must
pass through the solar position naturally creates the parabolic lower
boundary.  The black line shows the locus of the ILR, with the
scattering vectors (shown in green for this $m=3$ pattern in
Figure~\ref{fig.Jacobi}) being responsible for its slight misalignment
with the scattering tongue, which clearly shows the expected heating.
Similar features are seen in the \Gaia\ data
(Figure~\ref{fig.Gaia_all}) and have also been reported in simulations
\citep{Se12, SC14}.

The bottom panel of Figure~\ref{fig.spiral_ILR} shows $J_R$ as a
function of its conjugate phase $w_R$, which reveals that the density
in this space is also no longer smooth.  Again we see that the DF is
undisturbed over most of this projection of phase space.  The
variation at high $J_R$ is because stars of different $w_R$ that are
now at the Sun have different home radii, where the DF has different
normalizations.  Also the sharp features near $|w_R| \simeq 2$ are for
stars that were strongly scattered by the ILR and are now moving in
their epicycles, both inward and outward, past the Sun.

In an earlier paper, \citet{Se10} argued that all strongly scattered
stars should have the same values of $m w_\phi + l w_R$.  It is clear
that this expectation was wrong, since we now see two sharp features
near $|w_R| \simeq 2$.  The fact that it is symmetrical about $w_R=0$
(apocentre) is not a coincidence, since recalculating with a shift of
the intial phase of the spiral did not alter the symmetry about
$w_R=0$ or the positions of the scattering features, and only minor
details of the sharp features differed.

\begin{figure}
\begin{center}
\includegraphics[width=.98\hsize,angle=0]{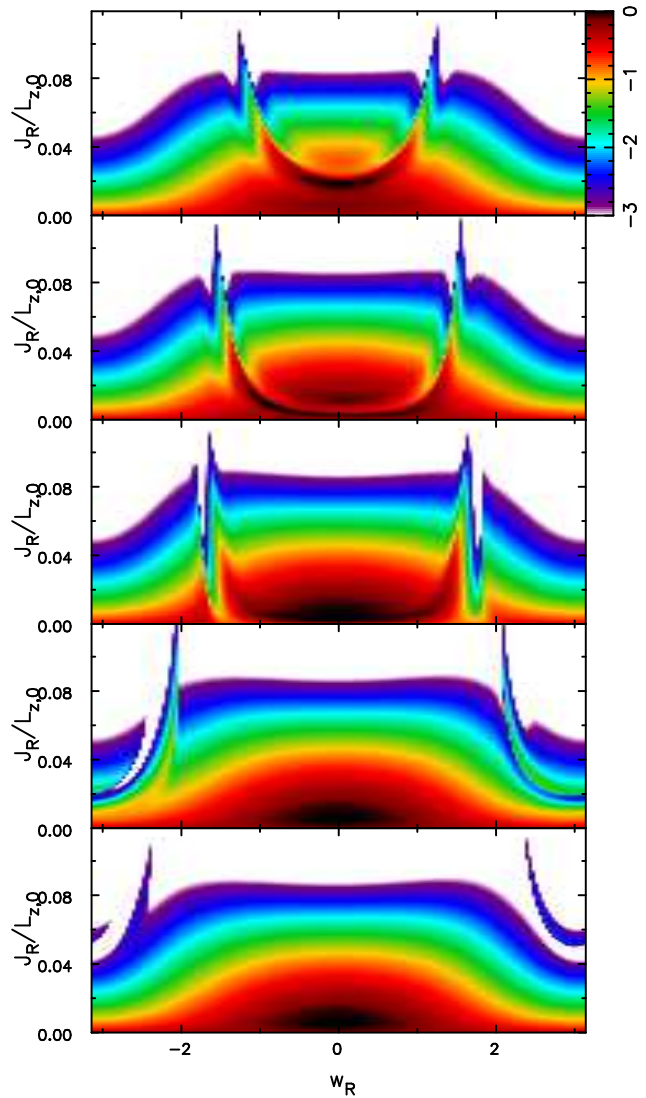} 
\end{center}
\caption{The effect on the $(w_R,J_R)$ distribution of changing the
  pattern speed of the spiral perturbation.  From top to bottom, the
  adopted pattern speeds in km~s$^{-1}$~kpc$^{-1}$ of our $m=3$
  pertubation are 17.1, 15.7, 14.8, 12.5, and 10.9.}
\label{fig.wrJr_ILR}
\end{figure}

\begin{figure}
\includegraphics[width=\hsize]{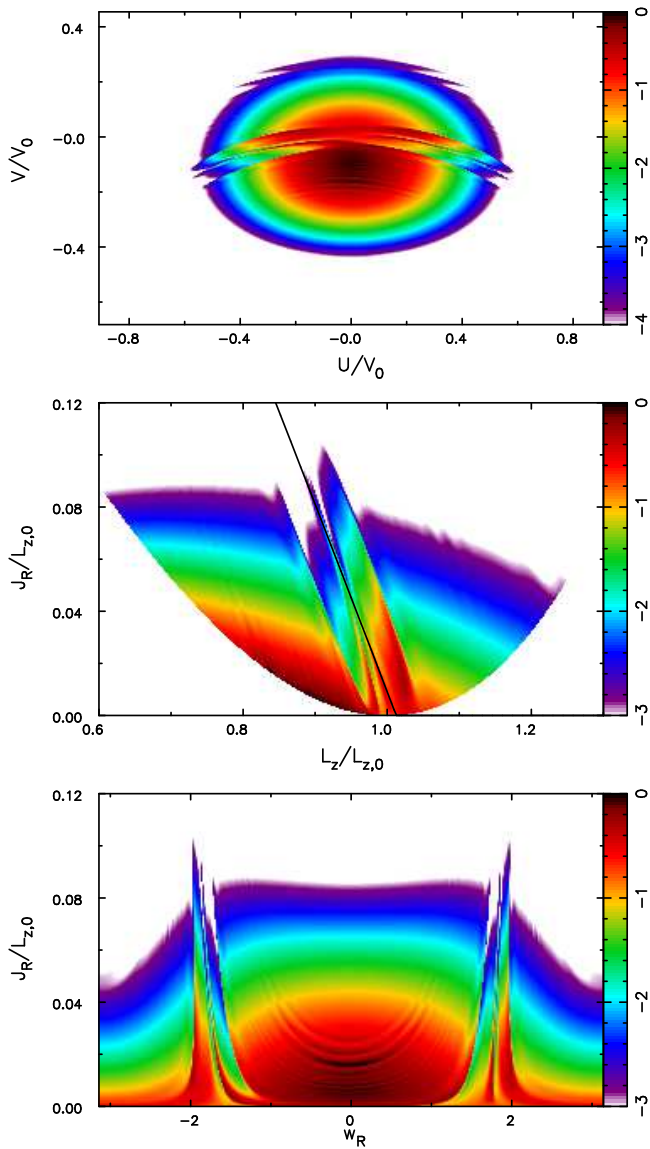} 
\caption{The predicted phase-space density of stars when the pattern
  speed of the spiral mode is raised so as to place the observer near
  the OLR.  The locus of the OLR is marked by the line in the middle
  panel, where resonance scattering has created a broad feature in
  which stars have been shifted both up and to the right and down and
  to the left.  Remarkably, the changes in the bottom panel resemble
  those in Figure~\ref{fig.spiral_ILR}.  The color scale has the same
  meaning as in Figure~\ref{fig.spiral_ILR}}
\label{fig.spiral_OLR}
\end{figure}

\begin{figure}
\includegraphics[width=\hsize]{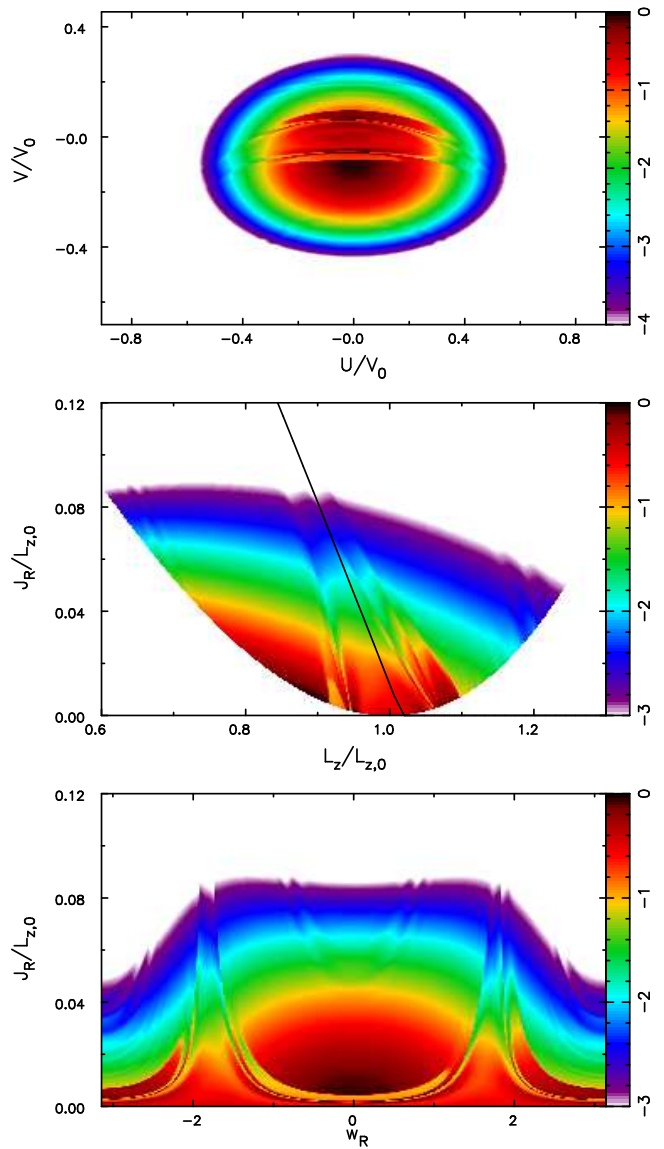}  
\caption{The predicted phase-space density of stars when the pattern
  speed of the spiral mode is adjusted so as to place the observer
  near CR.  The locus of the CR is marked by the line in the middle
  panel.  Changes to DF were remarkably mild, even though the
  perturbation amplitude was a maximum at this radius.  The color
  scale has the same meaning as in Figure~\ref{fig.spiral_ILR}}
\label{fig.spiral_CR}
\end{figure}

Changes to the pattern speed of the spiral cause, as expected, a
vertical shift to the scattering feature in the top panel of
Figure~\ref{fig.spiral_ILR} and a horizontal shift to the scattering
tongue in the middle panel.  The changes in the $(w_R,J_R)$
distribution are more interesting and are illustrated in
Figure~\ref{fig.wrJr_ILR}.  In each case, the observer remains at the
Sun's position.  The middle panel shows the effect of moving the ILR
to a point just interior to the Sun, which caused the stars near
apocentre $(w_R = 0)$ to be affected and again those with enough
radial action to have been strongly perturbed by the resonance
produced similar peaks, also near $|w_R|=\pi/2$ but curving towards
$w_R=0$ for lower values of $J_R$.  Raising the pattern speed to
higher values shifted the ILR farther inwards, giving rise to the
features shown in the top two panels of Figure~\ref{fig.wrJr_ILR}.  As
the ILR was shifted to larger radii (bottom two panels), the
scattering features very gradually moved towards $w_R=\pm\pi$ (for
stars at pericentre) and become weaker.  Note that the farther the
resonance from the solar radius in either direction, the less the DF
is disturbed for small $J_R$ over all $w_R$.  All this behaviour seems
physically very reasonable.

The consequence of scattering at the OLR is shown in
Figure~\ref{fig.spiral_OLR}, for which we adopted a pattern speed of
$\Omega_p = 40\;$km~s$^{-1}$~kpc$^{-1}$.  In this case, the observer
is just interior to the resonance, so the features in the bottom panel
curve towards $w_R=0$, since stars of low $J_R$ that have been
affected by the resonance are near their apocentres, which is the
opposite of the situation shown in Figure~\ref{fig.spiral_ILR}.  More
interesting is the feature in the middle panel; again the overall
slope of the feature is negative, but at this resonance the scattering
vectors have positive slope (Figure~\ref{fig.Jacobi}) with the
consequence that scattering shifts higher phase space density both up
and to the right and down and to the left.  Furthermore, since stars
are moved off resonance as they are scattered, we do not see the
pronounced peak that was created at the ILR
(Figure~\ref{fig.spiral_ILR}) and the spikes in bottom panel are
less pronounced.

It is noteworthy that adjusting the pattern speed to $\Omega_p =
27\;$km~s$^{-1}$~kpc$^{-1}$, so that the CR was at $8.13~$~kpc,
\ie\ just exterior to the Sun left phase space near the Sun almost
unchanged by the perturbation, as shown in Figure~\ref{fig.spiral_CR}.
This result is consistent with the prediction of formula
(\ref{eq.actions}), where stars simply change places with no heating.
Furthermore, \citet{LBK} showed that stars away from the major
resonances suffer no lasting changes to their integrals, and we find
similarly mild changes for observers located in a broad swath of radii
around CR where few stars are found that have been affected by either
of the Lindblad resonances.

\subsection{A material arm model}
\label{sec.materialarm}
A number of authors \citep[\eg][]{Gran12a, Gran12b, Baba13, Roca13,
  MK18} have argued that the spirals in their simulations are
swing-amplified features having pitch angles that change continuously
with time at a rate that is consistent with the local shear rate in
their adopted galaxy model.  In other words, the features are material
arms, and not density waves.  \citet{Hu18}, using the \texttt{galpy}
code \citep{Bo15}, adopted a potential of such a disturbance, combined
with a model of the MW bar, to calculate its effect on the
distribution of stars in $(U,V)$-space, and argued that some cases
resembled the distribution of \Gaia\ stars.

\begin{figure}
\includegraphics[width=\hsize]{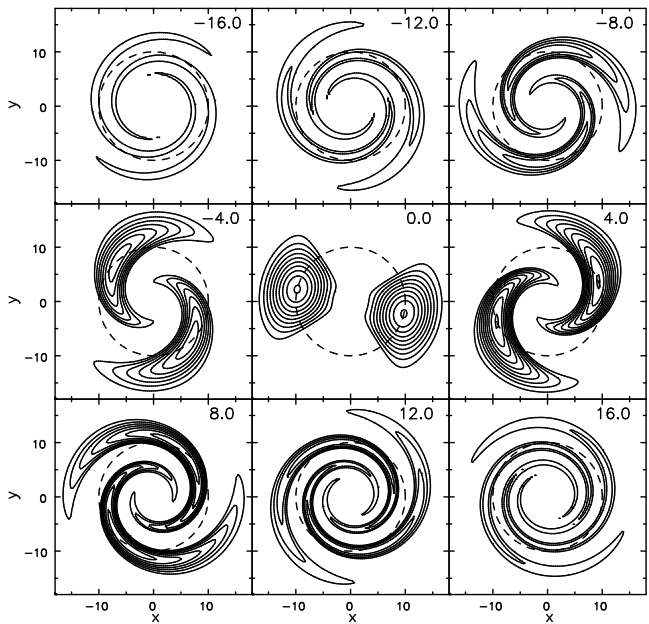}  
\caption{Contours of the negative part of the perturbation potential,
  which varies sinusoidally with azimuth, adopted for the material
  spiral arm model in a clockwise rotating galaxy model.  The model
  was motivated by the ``dust-to-ashes'' figure in \citet{To81}.  The
  times in the upper right corner of each frame are in units of
  $R_c/V_0$, where $R_c = 9\;$kpc is the radius of the potential
  maximum, which is marked by the dashed circle.  See the text for
  further details.}
\label{fig.swiphi}
\end{figure}

Although the disturbance they adopted indeed sheared, it did not seem
to us to resemble the classic image of swing-amplification presented
by \citet{To81}.  We have therefore adopted the different potential
perturbation that is illustrated in Figure~\ref{fig.swiphi}.  The
radial variation has peak amplitude at some chosen $R_c =
V_0/\Omega_c$ and is again as a quartic function that drops to zero at
$R = R_c(1 \pm 0.76)$.  The azimuthal behaviour is that of a
logarithmic spiral at all times, with the angle to the radius vector
$\psi = \Omega_c(t-t_0)$, which therefore changes as the perturbation
shears with the flow; it is radial at $t=t_0$ and $\psi>0$ for a
trailing spiral.  Its overall amplitude also scales with time as ${\rm
  sech}^2\beta$ where $\beta = 0.1[\Omega_c(t -t_0) - m]$, and $m$ is
the angular periodicity of the pattern; this shift in the argument
ensures that the peak amplitude occurs at a trailing pitch angle of
$45^\circ$, as expected for a flat rotation curve.  The factor 0.1
gives about the right variation in amplitude around the peak and
asymptotic growth and decay rates $\sim e^{-2\beta}$ for $|\beta| \gg
1$.  We choose the maximum potential amplitude $=0.02$ at $R_c
=9\;$kpc at a trailing angle of $\psi = 45^\circ$.

Note this potential perturbation was chosen to mimic a swing-amplified
spiral, but unlike that presented by \citet{To81}, it has no
underlying dynamics.  The potential function contains tightly-wrapped
ripples for $\Omega_c|t-t_0|\gg$ a few, which in reality should be
washed out by random motion in a warm disc.  Also, the disturbance in
our model has peak amplitude at $R_c$ at all times, which ignores the
radial propagation of the wave packet at the group velocity, as
occured in Toomre's dynamically self-consistent calculation.  However,
both these effects occur only when the spiral is quite tightly
wrapped, \ie\ when our perturbed amplitude is weak.  In summary, our
model captures the material shear of the disturbance while the
amplitude is large and the pattern open, which is the behaviour of the
simulations as described in the above-cited papers, while omitting any
dynamical details of the early and late evolution, when the amplitude
is very low.

\begin{figure}
\includegraphics[width=\hsize]{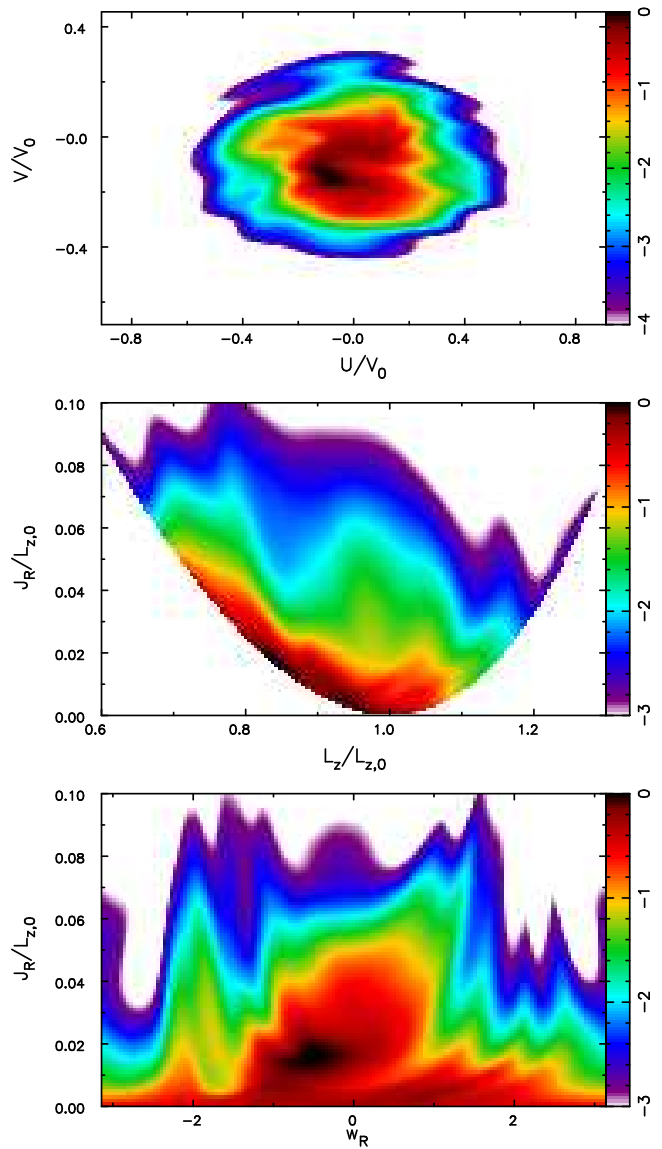} 
\caption{The distribution of stars in three different projections of
  phase space, observed from the Sun's location, that results from
  applying the bi-symmetric material spiral arm perturbation described
  in the text.  The top panel shows the distribution of velocities,
  the middle panel of actions, and the lower panel the angle $w_R$.
  The color scale has the same meaning as in
  Figure~\ref{fig.spiral_ILR}}
\label{fig.swingres}
\end{figure}

Figure~\ref{fig.swingres} presents three separate projections of the
phase space distribution of stars that results from this disturbance.
This quite strong perturbation has caused significant changes to the
DF; naturally, scaling down the amplitude of the perturbation results
in milder changes.  Unlike for the spiral mode considered above, the
individual features are quite broad and the entire distribution in
every projection has been sculpted by the perturbation.  We have
experimented with shifting the radius of the density maximum to
$R_c=7\;$kpc, \ie\ interior to the Sun's position, which again results
in changes to the entire distribution in all three projections, but
with the similarly broad peaks in differing locations.

We have also considered a 4-armed material spiral perturbations, which
resulted in more regular ``corrugations'' in all three projections of
the phase space density.  The corrugations were present both when the
potential maximum was interior or exterior to the Sun's location.

\subsection{A dressed mass clump}
\label{sec.massclump}
\citet{TK91} developed the idea that co-orbiting mass clumps within
the disc create a ``kaleidoscope'' of transient spiral features, which
are caused by the collective response of the underlying disc to
density inhomogeneities.  The local simulations by these authors, in
which shot noise from particles themselves created the inhomgeneities,
were extended to global simulations by \citet{DOVH}.  The latter
authors found that a sprinkling of heavy particles produced evolving
multi-arm spiral patterns in their rather low mass disc.  They also
reported that non-linear effects could cause activity to persist after
a single driving clump was removed.

It should be noted that each heavy particle quickly becomes dressed by
an extensive trailing wake, whose mass far exceeds that of the imposed
mass, and that the wake orbits at the angular rate of the perturbing
source mass.  The spiral patterns that result are caused by the
superposition of these wakes, each of which has its own pattern
speed.

In order to calculate changes in phase space density that would be
predicted by a simplifed model of this type, we must use the
perturbing potential of a dressed point mass.  The response of a
locally stable stellar disc to a point mass orbiting at the circular
speed was originally calculated by \citet{JT66} in the well-known
``sheared sheet'' model that neglects curvature.  The $(R,\phi)$
coordinates in the disc become
\begin{equation}
x=R-R_c \qquad\hbox{and}\qquad y=R_c\left[\phi -\Omega_c(t-t_0)\right]
\label{eq.coords}
\end{equation}
in the sheared sheet, where $\Omega_c$ is the circular angular
frequency at $R_c$, the radius of the sheet centre, which is the
location of the perturbing mass and therefore the corotation radius
of the disturbance.

\begin{figure}
\includegraphics[width=\hsize]{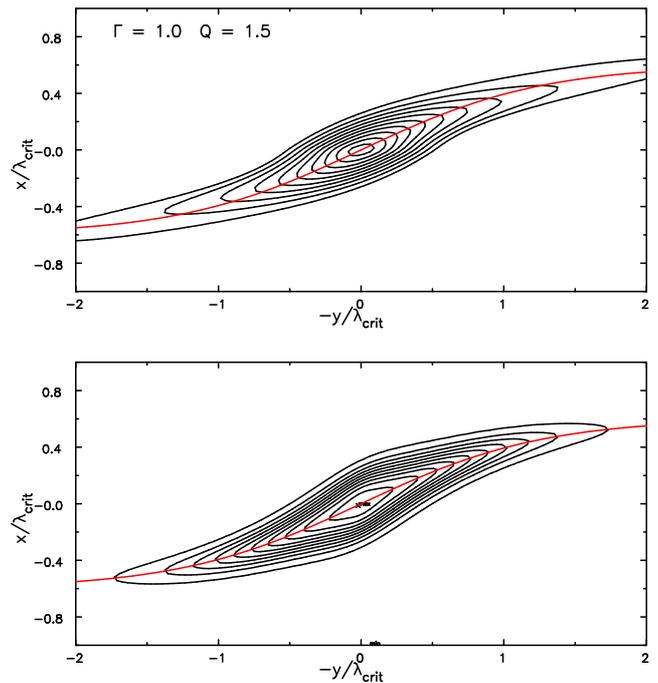} 
\caption{The upper panel contours the negative potential of the
  response density to a steady, coorbiting mass, computed by the method
  of \citet{JT66} in the sheared sheet, for a $V=$const disc with
  $Q=1.5$.  The $(y,x)$ coordinates, defined in the text, are in units
  of $\lambda_{\rm crit}$.  The lower panel shows the analytic
  approximation adopted here; the ridge-line of the approximate
  potential is marked in red and reproduced in the upper panel.}
\label{fig.JTwake}
\end{figure}

\begin{figure}
\includegraphics[width=\hsize]{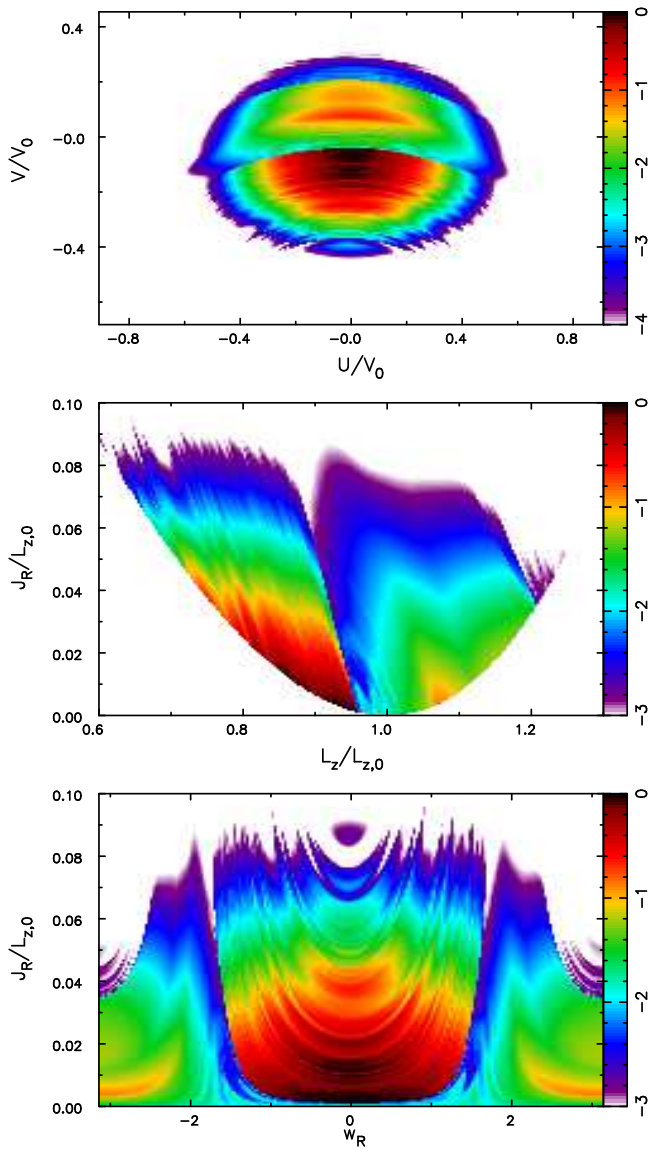} 
\caption{Three phase space projections resulting from a dressed mass
  clump model, whose centre orbits at $R=9\;$kpc, observed at
  $R=8\;$kpc, as usual.  The color scale has the same meaning as in
  Figure~\ref{fig.spiral_ILR}}
\label{fig.wake_model}
\end{figure}

The top panel of Figure~\ref{fig.JTwake} presents contours of the
perturbed potential computed using the mathematical apparatus devised
by \citet{JT66}.  The spatial scale of the response density in the
Figure can be converted to physical units by multiplying by
$\lambda_{\rm crit} = 4\pi G\Sigma/\kappa^2$, where $\Sigma$ is the
disc surface density, and $\kappa$ the epicyclic frequency, both
reckoned at $R=R_c$; in the self-similar Mestel disc $\lambda_{\rm
  crit} = 2\pi Rf$ at any radius, where $f$ is the fraction of active
mass.  The Figure illustrates the potential of a wake for a
$V=$const.\ disc with $Q=1.5$.

To use this perturbing potential in Dehnen's method, we must convert
$(R,\phi)$ to $(x,y)$ using equations (\ref{eq.coords}) and scale them
by an adopted value for $\lambda_{\rm crit}$.  It is inconvenient to
adopt this exact potential, because it is computed only over the
rectangle shown; the perturbed potential on the boundaries is small,
but non-zero, which would introduce mild discontinuities as stars
crossed the boundaries that would be problematic when integrating
orbits.  We have therefore adopted the analytic approximation for the
perturbing potential illustrated in the lower panel of
Figure~\ref{fig.JTwake}.  Its functonal form is
\begin{eqnarray}
\Phi_w(y,x) & = & -A e^{-a} \nonumber \\
a & = & y^{\prime2} + x^{\prime2} \left( 1 + 100 \sqrt{x^{\prime2} + y^{\prime2}} \right) \\
y^\prime & = & y / \lambda_{\rm crit} \nonumber \\
x^\prime & = & x / \lambda_{\rm crit} - 0.6 \tanh( \pi y^\prime/4 ),  \nonumber
\label{eq.approx}
\end{eqnarray}
with $A$ being a scaling constant.  The line $x = 0.6\lambda_{\rm
  crit} \tanh( 0.25\pi y/\lambda_{\rm crit})$, is the ridge line of
the approximate potential, and is marked in red in both panels of the
Figure.  We consider this function to be an adequate approximation to
the potential of a dressed particle.

It should be noted that the potentials shown in
Figure~\ref{fig.JTwake} are that of the steady response to a constant
perturbing mass.  \citet{JT66} reported that the wake grows quickly
after the introduction of a perturbing mass, and it takes only $\sim
5$ epicycle periods to asymptote to a steady response.  Thus the wake
of a growing mass would be only slightly weaker, unless the growth
rate is very high.

We chose $\lambda_{\rm crit} = \pi R_c/2$, appropriate for a 1/4 mass
Mestel disc for which half the circular speed arises from the disc
attraction, and which very roughly corresponds to the situation over
the massive part of the disc in the simulations by \citet{DOVH}.  We
also adopt the same overall $\hbox{sech}^2$ time dependence as in our
other models, set the radius of the co-orbiting mass clump to be at
$R_c=9\;$kpc, and the potential perturbation scale $A=0.02$ so as to
produce changes to the phase space density that are comparable in
magnitude to those presented previously.  The outcome is shown in
Figure~\ref{fig.wake_model}.  This perturbation produces significant
fine-scale substructure, especially for $L_z<L_{z,0}$, which also
appears in the lower panel for $|w_R| \la \pi/2$.  Shifting the
perturber to $R_c=7\;$kpc, \ie\ inside the solar radius, produced
similar fine substructure except that, in this case, it appeared for
$L_z>L_{z,0}$ and $|w_R| \ga \pi/2$.

The extensive fine-scale structure in this model differs from results
obtained for the two previous perturbations: the spiral mode model
(Figure~\ref{fig.spiral_ILR}) and the material arm model
(Figure~\ref{fig.swingres}).  In the present case, the wake has no
rotational symmetry, and a relatively narrow azimuthal extent.  A
sectoral harmonic ($e^{im\phi}$) decomposition of such a disturbance
would result in many $m$ components having significant amplitude, and
$l=\pm1$ resonances must arise for each separate $m$ component, which
are ever closer to the radius of the source as $m$ rises.  It seems
possible that the large number of fine features could have been
created by such resonances, but we defer detailed pursuit of this idea
to a later paper.

\subsection{Quasi-steady density waves}
\label{sec.qssh}
We do not present a calculation for the spiral density wave model of
\citet{BL96}.  Although \citet{Se11} demonstrated that their model had
serious problems, this criticism was ignored by \citet{Sh16}, who
continued to argue for the quasi-steady mode theory in his review.

Their model attributes ``grand design'' spirals to slowly growing,
bi-symmetric modes.  Since their picture also suggests that the mode
is ``quasi-steady'', it should not have begun to decay, as we assumed
for the transient mode model in \S\ref{sec.mode}.  Predictions for a
currently existing pattern could readily be calculated, indeed
\citet{WD00} devised his method to make predictions for the Milky Way
bar.  The most recent density-wave model for the Milky Way is that
presented by \citet{LYS}, which we could test.  However, \citet{Sh16}
argues that the situation in the Milky Way is now known to be more
complex than the earlier paper assumed, with evidence for a 4-arm
pattern, which he argues may possibly result from subharmonics or from
superposed patterns, perhaps with one being a response to the bar.

In the absence of a specific, testable model from this group, we make
a few general observations.  Their picture of spiral wave generation
invokes a $Q$-barrier to shield the mode from damping at the ILR, and
therefore the principal cause of features in the phase space
distribution in the transient spiral mode model would be absent in
their picture.  Also the CR has a neutral effect
(Figure~\ref{fig.spiral_CR}).  Furthermore, since spirals in this
model are long-lived and slowly evolving, these authors argue that
most galaxies will have supported rather few such patterns.  We would
therefore expect that the extensive structure observed by \Gaia\ in
the local phase space distribution of stars in the Milky Way
(Figure~\ref{fig.Gaia_all}) would not be predicted by their
quasi-steady spiral wave theory.

\subsection{Discussion}
\label{sec.discuss}
The results presented for the three separate transient spiral arm
models show that each gives rise to distinct predictions for the
change in the density of stars in all three projections of phase
space.  Furthermore, all three scenarios expect that the Galaxy has
supported many spiral episodes that have peaked at different radii,
which in the case of the \citet{SC14} mode theory would correspond to
different pattern speeds.

For many reasons, we do not here attempt to make a quantitative
comparison between the prediction of these models and the \Gaia\ data
presented in Figure~\ref{fig.Gaia_all}.  First, the data suggest that
the features in phase space result from a succesion of perturbations,
but we have so far computed the consequences of only single
disturbances.  Although scattering by a single spiral mode produces
only symmetric features in the $(w_R,J_R)$ plane, as shown in the
bottom panels of Figures~\ref{fig.spiral_ILR}--\ref{fig.spiral_OLR},
it may be possible that successive spiral disturbances could give rise
to some of the asymmetries in the third panel of
Figure~\ref{fig.Gaia_all}.  Second, it is likely that the local
phase-space distribution of stars has also been scuplted by other
perturbations, notably the bar in the Milky Way \citep{BHG} but also
perhaps tidally induced responses, which would need to be included in
any comprehensive model to confront the data.  Third, we have made no
attempt to match DF of the Milky Way, so that our choice of the
undisturbed DF yields a good match of the pertubed data to the
observations.  Fourth, features in phase space may be blurred over
time as stars are scattered by giant molecular clouds.  Building a
comprehensive model that includes all these aspects would be a major
undertaking, which we defer to later work.

Both the material arm model, Figure~\ref{fig.swingres}, and the
dressed clump model, Figure~\ref{fig.wake_model}, predict multiple
features from a single disturbance of large enough amplitude, and
therefore multiple such disturbances would probably produce quite
complex structure in all phase-space projections.  However, the
salient aspect of the data in the second panel of
Figure~\ref{fig.Gaia_all} are a few features of negative slope,
consistent with resonance scattering by a few patterns of low $m$.

The groups of stars from the \Gaia\ data in Figure~\ref{fig.mult_res}
that we highlighted in different colours were each selected to lie
within $\Delta L_z=0.01$ of a line of slope $-0.5$ whose intercept on
the $L_z$ axis was chosen by eye so that each selected group of stars
included one of the more prominent features in the second panel of
Figure~\ref{fig.Gaia_all}.  Because the loci of all the major
resonances have approximately this slope, we are unable to say whether
any selected feature corresponds to an ILR or an OLR, neither can we
identify the rotational symmetry of the perturbation that caused it,
and therefore we cannot estimate a pattern speed for any of these
possible disturbances.

The coloured stars in Figure~\ref{fig.mult_res} do not correspond to
every feature in the distribution of \Gaia\ stars, neither should
they.  The different scattering events, if that is what are
highlighted, will have occurred successively, the Milky Way hosts a
bar \citep{BHG}, and the disc has possibly been subject to other
perturbations that may have produced additional features in these
different projections.  However, the distributions of the variously
coloured stars in Figure~\ref{fig.mult_res} are quite similar to the
features in Figure~\ref{fig.spiral_ILR}, and the resemblance seems
closer than to those predicted by either of the other models.  Indeed,
we chose the pattern speeds in Figure~\ref{fig.wrJr_ILR} to be such
that the ILR of an $m=3$ spiral mode would correspond to each of the
highlighted features in Figure~\ref{fig.mult_res}.  Note that these
calculations were made for each pattern speed separately, whereas the
\Gaia\ data would reflect successive patterns that may account for
some of the asymmetries in the $(w_R,J_R)$-plane.  It should also be
cautioned that this superficially compelling comparison, particularly
in the $(w_R,J_R)$-plane, owes a lot to the fact that the appearance
in each projection is determined by the mapping of the 4D distribution
to each 2D surface, and the only really significant features are the
resonant ridges in the middle panel.

We conclude that the data offer little support for the dressed mass
clump model, none at all for the quasi-steady density wave model and,
while the material arm model may not be excluded, the sharper features
in the \Gaia\ data appear more consistent with the transient spiral
mode model of \citet{SC14}.

\subsection{Material arms \vs\ modes}
\label{sec.ruleout}
The popular claim \citep{Gran12a, Gran12b, Baba13, Roca13, MK18} that
spiral arms are material features that shear with the flow reflects
the apparent behaviour in the simulations.  \citet{SC84} at first
presented evidence that their spirals were shearing, swing-amplified
features, in agreement with these recent reports.  But a few years
later \citep[\eg][]{Se89} they showed that the density variations in
the {\em same simulations} could be decomposed into a number of
underlying coherent, steadily-rotating waves.

In order to understand how the two interpretations can give rise to
the same behaviour, the reader may find it helpful to watch the
animation at {\tt
  http://www.physics.rutgers.edu/$\sim$sellwood/spirals.html}, which
shows that an apparently shearing, swing-amplified spiral can result from
the super-position of two rigidly rotating patterns.  Note also that
the ``dust-to-ashes'' figure presented in \citet{To81} was {\em
  calculated} as the superposition of multiple steady responses to a
set of perturbers having a range of pattern speeds (private
communication, A. Toomre c1986).

Thus the superposition of multiple steadily-rotating patterns of
differing rotational symmetries, whose amplitudes vary on the
time-scale of a few orbits, can readily be imagined as giving rise to
the untidy and apparently random shearing spirals that are visible in
all simulations.  Indeed, \citet{SC14} explicitly reported their own
large-$N$ simulations of a low-mass disc that appeared to manifest the
kind of material arms observed in simulations by other groups and
showed from spectral analysis (their Fig 9) that the shearing features
resulted from the superposition of multiple coherent waves.

While the swing-amplified interpretation of simulated spirals is very
beguiling, and consistent with Toomre's picture, it begs the question
of what is the origin of the leading signal that is swing-amplified?
None of the above-cited papers that have argued for this
interpretation has even addressed this question, let alone provided a
satisfactory answer.

If the patterns were linear responses to shot noise, even of dressed
particles \citep{TK91}, their amplitude clearly should decrease as
$N^{-1/2}$.  This possibility was already disproved by \citet{SC14}
who compared the behaviour from simulations in which the particle
number ranged over three orders of magnitude.

Other tests by \citet{SC14} also ruled out that non-linear coupling
was needed to produce large-amplitude spirals.  After erasing any
coherent mass clumps by azimuthal shuffling of the particles in a
partly evolved simulation, they showed that the same coherent waves
were present in parallel simulations both after shuffling and in the
continued simulation without shuffling.  Had density variations at the
time of reshuffling been responsible for subsequent patterns, then
shuffling would have been equivalent to a fresh start, and the
amplitude of spirals would have grown as slowly, or more slowly
because random motion had increased slightly.  Instead they observed
more rapid and coherent growth than in the first start.  This result
proved that the axisymmetric changes caused by the earlier evolution
had created conditions for a vigorous, global, {\it linear\/}
instability that was not present in the original smooth disc.

The prominent self-excited spirals that develop in every large-$N$
simulation of disc galaxy models, that exclude forcing by clumps,
bars, or companions, require that they are true instabilities of the
dynamical system that is represented by the particles.  The
swing-amplified material arm interpretation, while superfically
attractive and a correct description of the apparent behavior, is
simply not a viable theory for spiral arm formation.  The
interpretation that the apparent features result from superposition of
a number of transient spiral modes was placed on a sound footing by
\citet{SC14} who also offered a mechanism that could produce
exponentially growing, global modes in a dynamically modified disc.
Our present finding that the distribution of local stars in
action-angle space is more consistent with multiple spiral modes than
with transient material disturbances provides further evidence in
support of their picture.

It might be objected that the transient disturbance adopted in
\S\ref{sec.materialarm} was not dynamically self-consistent, and
therefore not a fair test.  However, a fair test would be to consider,
as did \citet{To81}, a spiral that is the superposition of a number of
steady waves, which we have already shown would be more consistent
with the \Gaia\ data.

\section{Conclusions}
The second data release from the \Gaia\ mission has refined our view
of the phase-space distribution of stars near the Sun.  The components
of their motion parallel to the Galactic plane have long been known to
manifest detailed substructure, which was seen more sharply in the
first view of the new data \citep{Gaia2}.  Adopting a simple
axisymmetric model for the Milky Way that has a locally flat rotation
curve, enabled us to compute action-angle variables for a sub-sample
of likely thin-disk stars within a cylinder of radius 200~pc centered
on the Sun, and Figure~\ref{fig.Gaia_all} shows the density of stars
in four different projections of this 4D phase space.  The well-known
substructure in velocity space maps into rich substructure in these
other variables, which we note has some of the characteristic features
of Lindblad resonance scattering by multiple perturbations having a
range of pattern speeds and low-order rotational symmetry.

In order to determine whether some of these features might have been
caused by past spiral activity in the disc of the Milky Way, we
computed the changes to the distribution of stars in phase space that
would be caused by a single spiral pattern of the kind favored by each
of the following three current theories for spiral arm formation: the
transient spiral mode model \citep{SC14}, the swing-amplified model of
material arms \citep[\eg][]{Gran12a, Gran12b, Baba13, Roca13, MK18},
and the dresed mass clump model \citep{TK91, DOVH}.  We also argued
that the quasi-steady density wave model \citep{BL96} should not
create pronounced features in the phase-space distribution of stars
near the Sun.  For each of the first three cases, we adopted a
simplified transient potential that approximates that of an idealized,
isolated spiral.  We found that the inner Lindblad resonances of a
single pure transient spiral mode produced the narrow features in
phase space shown in Figure~\ref{fig.spiral_ILR}.  Multiple broad
features, see Figure~\ref{fig.swingres}, were created by a single
shearing spiral, or material arm, and multiple fine features,
Figure~\ref{fig.wake_model}, resulted from the wake of single dressed
particle.  For each model, we experimented with changes to the pattern
speed or, in the case of material arms, the radius of the perturbed
potential minimum, and found that the locations of the features in
each projection varied in ways that made dynamical sense.

We argued that the \Gaia\ data are inconsistent with the dressed
particle model, and while they could be consistent with the material
arm model, some of the narrower scattering features in
Figure~\ref{fig.Gaia_all} more strongly resemble the prediction of
spiral modes that are expected to have had a variety of pattern speeds
and rotational symmetries.  Other considerations discussed in
\S\ref{sec.ruleout} strongly disfavour the swing-amplified material
arm model, although we also explain that the superposition of a number
of steady spiral modes can give the visual impression swing-amplified
shearing evolution.

We do not here claim to have presented a model that can account for
all the features in the 4D phase space distribution.  The rich
substructure in the distribution of \Gaia\ stars deserves additional
effort to fully understand its origin.

\section*{Acknowledgements}
We thank Scott Tremaine and James Binney for discussions and an
anonymous referee for suggestions that have improved the presentation.
JAS acknowledges the hospitality of Steward Observatory where most of
this work was done.  This work has made use of data from the European
Space Agency (ESA) mission
\Gaia\ (\url{https://www.cosmos.esa.int/gaia}), processed by the
\Gaia\ Data Processing and Analysis Consortium (DPAC
\url{https://www.cosmos.esa.int/web/gaia/dpac/consortium}).  Funding
for the DPAC has been provided by national institutions, in particular
the institutions participating in the \Gaia\ Multilateral Agreement.


\bsp	
\label{lastpage}
\end{document}